\documentstyle[preprint,aps,eqsecnum,epsf,floats]{revtex}
\tighten

\begin{document}
\newcommand{\dencas}{\mbox{$m_\Lambda - m_\Xi$}}
\newcommand{\denn}{\mbox{$m_\Lambda - m_N$}}
\newcommand{\hf}{\mbox{$h_F$}}
\newcommand{\hd}{\mbox{$h_D$}}
\newcommand{\hc}{\mbox{$h_C$}}
\newcommand{\jay}[2]{\mbox{${\cal J}(\Delta m^{#1}_{#2})$}}
\newcommand{\h}{\mbox{${\cal H}$}}
\newcommand{\kay}[3]{\mbox{${\cal K}(#1,\Delta m^{#2}_{#3})$}}
\newcommand{\gee}[3]{\mbox{$G_{#1,#2,#3}$}}
\newcommand{\geetwid}[3]{\mbox{$\tilde G_{#1,#2,#3}$}}
\newcommand{\be}{\begin{eqnarray}}
\newcommand{\ee}{\end{eqnarray}}
\newcommand{\front}{\mbox{$ 
{m_K^2  \over 16\pi^2 f_K^2}\ln\left({m_K^2 \over\Lambda_\chi^2}\right)$}}

{\tighten
\preprint{\vbox{
\hbox{DUKE-97-152}
\hbox{DOE--ER/41014-28-N97}
}}

\title{ The Weak $\Lambda \Lambda K^0$ Interaction}
\author{Martin  J. Savage}
\address{
Department of Physics, University of Washington, Seattle, WA 98915 
\\ {\tt savage@phys.washington.edu}}
\author{Roxanne P. Springer}
\address{Duke University Department of Physics, Durham, NC 27708
\\ {\tt rps@phy.duke.edu}}
\maketitle 
\begin{abstract}
We calculate the leading contribution to the 
weak $\Lambda \Lambda K$ coupling
in heavy baryon chiral perturbation theory, including
the SU(3) breaking terms arising at one-loop. 
This coupling gives the leading one-boson exchange 
amplitude for the process 
$\Lambda\Lambda\rightarrow\Lambda n$ and is an important component 
in understanding
the decay of $\Lambda\Lambda$-hypernuclei.
We find the S-wave and P-wave amplitudes to be stable under
one-loop corrections, but the associated errors become large.
We discuss the theoretical errors and potential effects
of higher order terms.
\end{abstract}

\section{Introduction}

Experiments at Brookhaven (BNL-AGS-E885, E906) and
the Japanese Hadron Facility (KEK-E373)
are planned to measure 
the properties and decays of nuclei containing more than
one strange baryon, such as  $\Lambda\Lambda$-hypernuclei.
Some observations already exist \cite{exp}.
In addition to the conventional decay mechanisms that exist for
singly strange hypernuclei; mesonic decay such as
$\Lambda\rightarrow N\pi$ (suppressed in heavy nuclei)
and weak scattering such as $\Lambda N\rightarrow NN$, there is a decay
amplitude for  $\Lambda\Lambda$-hypernuclei
arising from $\Lambda\Lambda\rightarrow \Lambda N$.
Unlike $\Lambda N$ weak scattering there is no contribution from single $\pi$
exchange to $\Lambda\Lambda$ weak scattering
because of the absence of a strong $\Lambda\Lambda\pi$ coupling.
The leading amplitude therefore comes from the exchange of a virtual $K^0$.
The strong $\Lambda N K^0$ interaction can be determined from the 
semi-leptonic decay $\Lambda\rightarrow pe^-\overline{\nu}_e$, using PCAC.
However, the nonleptonic weak $\Lambda\Lambda K^0$ vertex cannot be measured
directly.
The approximate SU(3) flavor symmetry of
the strong interactions  can be used in combination with the 
assumption of octet-dominance in the nonleptonic weak interactions to relate
the $\Lambda\Lambda K^0$ vertex  to  amplitudes for nonleptonic hyperon
decays $B\rightarrow B^\prime\pi$.
In order to systematically  include SU(3) breaking contributions  to the decay
we use chiral perturbation theory.
Such calculations have been performed for the nonleptonic hyperon decays 
\cite{BSW85a,ej} and for the $NNK$ vertices \cite{knn}, assuming that
the breaking is dominated by terms of the form $m_s\log m_s$.
We will determine the  leading SU(3) breaking corrections of the
form  $m_s\log m_s$ to the $\Lambda\Lambda K^0$ couplings.
These couplings may then be used as an input into the nuclear codes
that presently exist (e.g., \cite{bennhold}) in order to determine the
modification to the decay of  $\Lambda\Lambda$-hypernuclei.

An additional motivation for studying the $\Lambda \Lambda K^0$
couplings  comes from  an observation that some consider to 
cast doubt on the convergence of  SU(3) chiral perturbation theory.
A long-standing problem for chiral perturbation theory
has been its failure to reliably 
predict the P-wave amplitudes in the nonleptonic
decays of octet baryons.  
The loop corrections are relatively large compared to the tree level 
values\cite{ej}
but this is thought to have more to do with the 
significant but accidental cancellation among
tree level terms than an instability in the chiral expansion \cite{georgi}.  
However, it is important to determine if the situation 
with the P-wave amplitudes in the octet decays is universal, 
and whether the predictions of chiral perturbation theory 
can be considered robust, with certain understood exceptions.  
To this end, we  calculated  the
$KNN$ couplings and found that the P-waves amplitudes 
are more stable against SU(3) breaking contributions
than the S-waves amplitudes, and that neither is as  badly 
behaved as P-wave amplitudes for octet baryon decay\cite{knn}.  
That work has been used \cite{bennhold} in an attempt to
understand the discrepancy between the  neutron induced versus proton induced
decay rates of $\Lambda$ hypernuclei. 
With the emergence of more accurate experimental results, 
we expect that the importance of the $KNN$ coupling for these processes
will be determined.
We hope that the future experimental studies 
of $\Lambda\Lambda$-hypernuclei
and the substantial theoretical efforts presently 
underway will establish the $\Lambda \Lambda K^0$ vertex.
Together with the data on singly-strange hypernuclei, the pattern of the weak
octet baryon interactions will become clearer.

\section{The Chiral Lagrangian for Nonleptonic Interactions}

This section closely follows the section in 
\cite{knn} where the Lagrangian and relevant notation
is presented.  Weak nonleptonic
hypernuclear decay takes place in an environment where
typical momentum transfers are smaller than the chiral
symmetry breaking scale; $p < \Lambda_\chi$, where
$\Lambda_\chi \sim 1$ GeV.
At these scales, it is appropriate to use as fundamental
fields the lowest mass octet and decuplet baryons and the 
octet of pseudo-Goldstone bosons.  The octet of baryons
is 
\begin{eqnarray}\label{octet}
B_v =
\pmatrix{ {1\over\sqrt2}\Sigma_v^0 + {1\over\sqrt6}\Lambda_v &
\Sigma_v^+ & p_v\cr \Sigma_v^-& -{1\over\sqrt2}\Sigma_v^0 +
{1\over\sqrt6}\Lambda_v&n_v\cr \Xi_v^- &\Xi_v^0 &- 
{2\over\sqrt6}\Lambda_v\cr }
\ \ \ ,
\end{eqnarray}

\noindent and the decuplet is 
\begin{eqnarray}\label{decuplet}
& & T^{111}_v  = \Delta^{++}_v, \ \ T^{112}_v = {1\over\sqrt{3}}\Delta^{+}_v, 
\ \ T^{122}_v = {1\over\sqrt{3}}\Delta^{0}_v, \ \ T^{222}_v = \Delta^{-}_v,
\nonumber\\
& & T^{113}_v = {1\over \sqrt{3}}\Sigma^{*+}_v,\ \ 
T^{123}_v  = {1\over\sqrt{6}}\Sigma^{*0}_v, \ \ 
T^{223}_v = {1\over\sqrt{3}}\Sigma^{*-}_v,
\ \ T^{133}_v = {1\over\sqrt{3}}\Xi^{*0}_v, 
\nonumber\\
& & T^{233}_v = {1\over\sqrt{3}}\Xi^{*-}_v, \ \ T^{333}_v = 
\Omega^-_v  \ \ \ .
\end{eqnarray}
The subscript $v$ labels the  four-velocity 
of the baryon.  The average mass of
the octet of baryons has been explicitly removed from
the Lagrangian \cite{MJ}.  The octet of mesons is contained in 
\begin{eqnarray}
\Sigma  = \xi^2= {\rm exp}\left( {2 i M\over f} \right) \ \ \ ,
\end{eqnarray}
where
\begin{eqnarray}
M = 
\left(\matrix{{1\over\sqrt{6}}\eta+{1\over\sqrt{2}}\pi^0&\pi^+&K^+\cr
\pi^-&{1\over\sqrt{6}}\eta-{1\over\sqrt{2}}\pi^0&K^0\cr
K^-&\overline{K}^0&-{2\over\sqrt{6}}\eta\cr}\right)
\ \ \ \  .
\end{eqnarray}
Using this notation, a vector and an axial vector current
can be defined:
\begin{eqnarray}
V_\mu&=&{1 \over 2} (\xi\partial_\mu\xi^\dagger + 
\xi^\dagger\partial_\mu\xi) 
\nonumber \\
A_\mu&=&{i \over 2} (\xi\partial_\mu\xi^\dagger - 
\xi^\dagger\partial_\mu\xi) 
\ \ \ .
\end{eqnarray}

With these elements, and the covariant chiral derivative
${\cal D_\mu}= \partial_\mu+[V_\mu,  \; ]$,
an $SU(3)_L\otimes SU(3)_R$ symmetric Lagrangian can be
formed.  The lowest dimension operators will dominate observables,
with the higher dimensions suppressed by increasing powers
of ${p \over \Lambda_\chi}$.  
To calculate the $\Lambda\Lambda K^0$ coupling to one loop, we need both
strong and weak interactions of the baryons with the mesons,
\begin{eqnarray}
{\cal L} & = & {\cal L}^{st} + {\cal L}^{wk} 
\ \ \ \  .
\end{eqnarray}
The strong interactions are described by 
\begin{eqnarray}\label{strongl}  
{\cal L}^{st} &=& i {\rm Tr} \bar B_v \left(v\cdot {\cal D} \right)B_v 
+ 2 D\  {\rm Tr} \bar B_v S_v^\mu \{ A_\mu, B_v \} 
+ 2 F\ {\rm Tr}  \bar B_v S_v^\mu [A_\mu, B_v] 
\nonumber \\
&&- i \bar T_v^{\mu} (v \cdot {\cal D}) \  T_{v \mu} 
+ \Delta m \bar T_v^{\mu} T_{v \mu} 
+ {\cal C} \left(\bar T_v^{\mu} A_{\mu} B_v + \bar B_v A_{\mu} T_v^{\mu}\right)
\nonumber\\ 
&& + 2 {\cal H}\  \bar T_v^{\mu} S_{v \nu} A^{\nu}  T_{v \mu} 
+ {f^2 \over 8} {\rm Tr} \partial_\mu \Sigma \partial^\mu \Sigma^\dagger 
+\ \cdots \ \ \ \  ,
\end{eqnarray}
where $f$ is the meson decay constant 
and only the lowest dimension
operators are retained in the expansion. 
The axial couplings $F, D, C$, and ${\cal H}$
which appear in ${\cal L}^{(st)}$ could in principle be
determined by matching this effective Lagrangian onto
the appropriate calculation in QCD.  Since such a computation in 
QCD  has not been performed, these parameters are taken as
unknowns and determined through a fit to experimental data.
The $\Delta s=1$ weak interactions required for 
this computation are described by
the Lagrange density (assuming octet dominance) 
\begin{eqnarray}\label{weakl}
{\cal L}^{wk} &=& 
G_Fm_\pi^2 f_\pi \Big( h_D {\rm Tr} {\overline B}_v 
\lbrace \xi^\dagger h\xi \, , B_v \rbrace \;  
+ \;  h_F {\rm Tr} {\overline B}_v  
{[\xi^\dagger h\xi \, , B_v ]} \; \nonumber \\ &&
+  h_C {\overline T}^\mu_v
(\xi^\dagger h\xi) T_{v \mu}  \; + \; 
h_\pi {f^2 \over 8} {\rm Tr} \left(  h \, \partial_\mu 
\Sigma \partial^\mu \Sigma^\dagger  \right) 
\ + \ \cdots\ \ \ \Big) \ \ \ \   ,
\end{eqnarray}
where
\begin{eqnarray}
h = \left(\matrix{0&0&0\cr 0&0&1\cr 0&0&0}\right)  \ \ \ 
\end{eqnarray}

This Lagrange density contains additional unknown couplings that must be
determined from data:
 $h_D, h_F, h_\pi$, and $h_C$.  
The pion decay constant is known to be $f_\pi \sim 132$ MeV. 
We have inserted factors of $ G_Fm_\pi^2 f_\pi $ in Eq.~\ref{weakl}\ 
so that the constants $h_D, h_F$, and $h_C$ are 
dimensionless.

\section{$\Lambda \Lambda K^0$ Amplitudes}

The amplitude for the weak $\Delta s = 1$  nonleptonic 
interaction $\Lambda\Lambda K^0$ is given by
\begin{eqnarray}\label{spamp}
{\cal A} = i G_F m_\pi^2\  {f_\pi \over f_K}\ \overline{\Lambda}_v \  
\left[    {\cal A}^{(S)}  +  2 {k \cdot S_v \over \Lambda_\chi} {\cal A}^{(P)}  
\right] \Lambda_v 
\, \, ,
\end{eqnarray}
where $k$ is the outgoing momentum of the $K^0$ and
\begin{eqnarray}
{\cal A}^{(L)} = {\cal A}_0^{(L)} + {\cal A}_1^{(L)} + \cdots \ ,
\end{eqnarray}
with $L$=S for the S-wave amplitude, or $L$=P for the P-wave
amplitude.  The subscripts in the above expansion for each $L$-wave 
indicate the order in chiral perturbation theory.

\subsection{S-Wave Vertices}

The tree level amplitude, as shown in Fig.~\ref{streefig} gives
\begin{eqnarray}\label{stree}
{\cal A}_0^{(S)} (\Lambda \Lambda K^0) &=& - h_D 
\ \ \ \ .
\end{eqnarray}
\noindent The loop corrections are shown in Fig.~\ref{sloopfig} and
evaluate to 
\begin{eqnarray}\label{sloop}
{\cal A}_1^{(S)} (\Lambda \Lambda K^0)& =& 
{m_K^2  \over 16\pi^2 f_K^2}\ln\left({m_K^2 \over \Lambda_\chi^2}\right)
\Big[{10 \over 3}h_D  
+h_D\left(-{11 \over 3}D^2-9F^2\right)+18h_FDF \Big] \nonumber \\
&& + h_C C^2 {\cal J} (\Delta m ) 
- {\cal A}_0^{(S)} (\Lambda \Lambda K^0){\cal Z}_\Psi
\end{eqnarray}
where the contribution from wavefunction 
renormalization (not shown in the figure) is given by 
\begin{eqnarray}
{\cal Z}_\Psi  = {m_K^2  \over 16\pi^2 f_K^2} \ln ({m_K^2 \over \Lambda_\chi^2}) 
\Big( 18F^2  +  {14 \over 3}D^2\Big) 
+  C^2  {\cal J}(\Delta m) 
\ \ \ \ \ ,
\end{eqnarray}
and the function ${\cal J}(\delta)$ is
\begin{eqnarray}
{\cal J}(\delta)  =  {1 \over 16\pi^2 f_K^2}
&&\left[(m_K^2  -2 \delta^2)\ln\left({m_K^2 \over  \Lambda_\chi^2}\right) 
+ 2\delta \sqrt{\delta^2-m_K^2 } 
\ln \left( { \delta-\sqrt{\delta^2-m_K^2+ i\epsilon}  \over
\delta+\sqrt{\delta^2-m_K^2 + i\epsilon}}\right) \right]
\ \ .
\end{eqnarray}
The argument of ${\cal J}$ is the
mass splitting between the decuplet and octet of baryons.
It is important to note that 
we have not included wavefunction renormalization for the $K^0$ 
so that double counting is avoided in the structure calculations\cite{bennhold} .

\subsection{P-Wave Vertices}

The tree level P-wave amplitude arises from two pole graphs, as shown in 
Fig.~\ref{ptreefig} , 
\begin{eqnarray}
{{\cal A}^{(P)}_0 (\Lambda \Lambda K^0) \over \Lambda_\chi}
 ={1 \over 6}{(D-3F)(3\hf\ -\hd\ ) \over \dencas\ }-
{1 \over 6}{(D+3F)(\hd+3\hf) \over \denn\ }
\ \ \ \  .
\end{eqnarray}

The loop level contributions are shown in Fig.~\ref{ploopfig}
and Fig.~\ref{ploop2}.
We  present the contributions from each graph separately.
As with the S-waves,  the wavefunction renormalization
on the external kaon is not included, 
since that will be accounted for in the structure calculations.
We find that the full amplitude is the sum of the following contributions:
\begin{eqnarray}
a & = & \front \left[ {D+3F \over \denn}
 \left({5 \over 12}\hd+{5 \over 4}\hf \right) \right.
\nonumber \\
&&\left.+ {D-3F \over \dencas} \left({5 \over 12}\hd-
{5 \over 4}\hf \right)\right] 
\end{eqnarray}
\begin{eqnarray}
b  & = & \front \left[{5 \over 36}{(3F+D)(\hd+3\hf) \over \denn} 
\right.\nonumber \\
&&\left. +{5 \over 36}{(3F-D)(3\hf-\hd) \over \dencas}\right] 
\ee \be
c & =& \front\left[{1 \over 6}{(\hd+3\hf)(D-3F) \over \denn}
\left({19 \over 9}D^2+{4 \over 3}DF-3F^2\right) \right.
\nonumber \\
 &&\left. +{1 \over 6}{(3\hf-\hd)(D+3F) \over \dencas}
\left(-{19 \over 9}D^2+{4 \over 3}DF+3F^2\right)\right] 
\ee \be
d & = &\jay{}{}C^2\h{5 \over 27}{\hd+3\hf \over \denn}
+\jay{}{}C^2\h{5 \over 27}{3\hf-\hd \over \dencas} 
\ee \be
e & = & {1 \over 9}\kay{m_K}{}{}
C^2{(\hd+3\hf)(D-3F) \over \denn} \nonumber \\
 &&-\kay{m_K}{}{}C^2{\hd+3\hf \over \denn}{4D \over 9} 
\nonumber \\
&& +{3\hf-\hd \over \dencas}C^2\left(-{1 \over 3}(D-F)\kay{m_K}{}{}
 -{2 \over 9}D\kay{m_\eta}{}{}\right) 
\ee \be
f &=&\front{(D+3F)(D-3F) \over \denn}
\left(-{2 \over 9}D(\hd+3\hf) \right. \nonumber \\
&&\left.  -{1 \over 12}
(D+3F)(\hd-3\hf)-{3 \over 4}(D-F)(\hd+\hf)\right) \nonumber \\
&&+\front{(D+3F)(D-3F) \over \dencas}
\left(-{1 \over 12}(D-3F)(\hd+3\hf)\right. \nonumber \\
&& \left. +{2 \over 9}D(3\hf-\hd)+{3 \over 4}(F+D)(\hf-\hd)\right) 
\ee \be
g &=&{1 \over 3}\jay{}{}C^2\hc{D+3F \over \denn}
\nonumber \\
 && +{1 \over 3}\jay{}{}C^2\hc{3F-D \over \dencas} 
\ee \be
h &=&-{1 \over 36} D h_\pi \front 
\ee \be
i &=& \front h_\pi D \left({7 \over 108}
(D-3F)^2-{1 \over 4}(D-F)(D+3F) \right .\nonumber \\
 &&  \left. -{ 1 \over 4}(F+D)(D-3F)\right)  
\ee \be
j & =& \jay{}{}C^2\h h_\pi {5 \over 18}
\ee \be
k &=&\gee{m_\pi}{m_K}{\Delta m}C^2 h_\pi(D-3F){1 \over 6} \nonumber \\
&&+\gee{m_\pi}{m_K}{\Delta m}C^2 h_\pi\left({D \over 6}-{F \over 2}\right)
-\gee{m_\eta}{m_K}{\Delta m}C^2 h_\pi{D \over 9}
\ee \be
l &=& h_\pi(D-3F)
\Big(\geetwid{m_K}{m_\eta}{0}({D^2 \over 16}-{1 \over 8}DF-{3 \over 16}
F^2) \nonumber \\
&&  - \geetwid{m_K}{m_\eta}{\Delta m}{1 \over 24}C ^2  \nonumber \\
&&+\geetwid{m_\pi}{m_K}{0}({9 \over 16}F^2-{9 \over 8}FD-{3 \over 16}D^2) 
\nonumber \\
&&   -\geetwid{m_\pi}{m_K}{\Delta m}{C^2 \over 4}  \nonumber \\
&&+{C^2 \Delta m \over \dencas}({1 \over 2} \gee{m_\pi}{m_k}{\Delta m} +
{1 \over 12} \gee{m_\eta}{m_k}{\Delta m}) \Big) \nonumber \\
&&+h_\pi(D+3F)
\Big(\geetwid{m_K}{m_\eta}{0}({D^2 \over 16}+{1 \over 8}FD-{3\over 16}F^2) 
   \nonumber \\
&& +\geetwid{m_\pi}{m_K}{0}(-{3 \over 16}D^2+{9 \over 8}FD+{9 \over 16}F^2)
\nonumber \\
&&+   \geetwid{m_\pi}{m_K}{\Delta m}{C^2 \over 8} \nonumber \\
&&-{1 \over 4}{C^2 \Delta m \over \denn} \gee{m_\pi}{m_k}{\Delta m} \Big) 
\ee  \be
\Psi_{off} &=&-{1 \over 6}{(3\hf-\hd)(D-3F) \over \dencas}
\Big[\front({17 \over 3}D^2+10DF+15F^2)+{13 \over 3}C^2\jay{}{}\Big]
\nonumber \\  && +{1 \over 6}{(\hd+3\hf)(D+3F) \over \denn}
\Big[\front({17 \over 3}D^2-10DF+15F^2)+C^2\jay{}{}\Big] 
\ee \be
\Psi &=&-{\cal A}^{(P)}_0(\Lambda \Lambda K^0){\cal Z}_\Psi
\end{eqnarray}

The external baryon wavefunction renormalization is contained
in $\Psi$, with the function ${\cal Z}_\Psi$ defined as before
in the S-wave formulas.
The additional functions which appear in the above expressions are given
by
\begin{eqnarray}
{\cal K}(m,\delta)  =  {1 \over 16\pi^2 f_K^2} & & \Big\{
(m^2 - {2\over 3}\delta^2) \ln({m^2 \over \Lambda_\chi^2}) 
\nonumber\\
& & +{2 \over 3}{1 \over \delta}\Big[(\delta^2-m^2)^{3/2}
\ln\left( {\delta-\sqrt{\delta^2-m^2+i\epsilon} \over 
\delta+\sqrt{\delta^2-m^2+i\epsilon}}\right)
+ \pi m^3 \Big]\Big\}
\ \ \ ,
\end{eqnarray}

\begin{eqnarray}
G_{m_1,m_2,\delta} & = & {m_1^2 \over m_1^2-m_2^2}
{\cal K}(m_1,\delta) + {m_2^2 \over m_2^2-m_1^2}
{\cal K}(m_2,\delta)\ \  , 
\nonumber \\
\tilde{G}_{m_1,m_2,\delta} & = & {m_1^2 \over m_1^2-m_2^2}
{\cal J}(m_1,\delta) + {m_2^2 \over m_2^2-m_1^2}
{\cal J}(m_2,\delta)
\ \ \ \ .
\end{eqnarray}
\noindent  As before, the argument $\delta$ is
the average decuplet-octet mass splitting.

\section{Numerical Results}

Our  numerical results 
for the $\Lambda\Lambda K^0$ couplings depend upon a 
number of parameters that must be fit to experimental data.  
For tree level estimates, we use
parameters obtained by fitting tree level expressions to the
experimentally measured rates and asymmetries in 
the semileptonic decays of  octet baryons, and
the S-wave amplitudes in the nonleptonic decay of 
octet  baryons. 
The systematic theoretical error associated with the
neglect of SU(3) breaking is included in the fit by
increasing the error bars of the experimental measurements
to $20\%$. 
We find that the  tree level expressions for 
the S- and P- wave amplitudes are 
\be
{\cal A}^{(S)}_0 (\Lambda \Lambda K^0) = 0.55 \pm 0.19 \nonumber \\
{{\cal A}^{(P)}_0 (\Lambda \Lambda K^0)\over \Lambda_\chi} = -5.2 \pm 1.0
\ \ \ ,
\ee
using the fitted values 
$D= 0.82 \pm 0.07\ ,  F= 0.48 \pm 0.06\ , h_D = -0.55 \pm 0.19 \ ,$ and 
$h_F = 1.37 \pm 0.11$.

Loop level numerical results for the $\Lambda\Lambda K^0$ couplings depend
upon parameters extracted at loop level from other processes.
We find that simultaneously fitting 
the semileptonic decays of the octet of baryons\cite{mj}, 
the strong decays of the decuplet of baryons\cite{bss}, 
and the S-wave nonleptonic decays of the octet of baryons\cite{ej} yields
for $\Delta m=0$
\be
{\cal A}^{(S)}_0(\Lambda \Lambda K^0) + 
{\cal A}^{(S)}_1(\Lambda \Lambda K^0) & = & 0.43 \pm 0.89 \nonumber \\
{\cal A}^{(P)}_0(\Lambda \Lambda K^0) +
{\cal A}^{(P)}_1(\Lambda \Lambda K^0) & = & -4.9 \pm 1.9 
\ \ \ \  .
\ee
This suggests that the S-wave predictions are probably unreliable in
chiral perturbation theory.  
When $\Delta m = 200$ MeV is used for the $\Lambda \Lambda K^0$ couplings,
we obtain
\be
{\cal A}^{(S)}_0(\Lambda \Lambda K^0) + 
{\cal A}^{(S)}_1(\Lambda \Lambda K^0) & = & 0.28 \pm 0.82 \nonumber \\
{\cal A}^{(P)}_0(\Lambda \Lambda K^0) +
{\cal A}^{(P)}_1(\Lambda \Lambda K^0) & = & -4.8 \pm 4.1 
\ \ \ \  .
\ee
Both the S-wave and P-wave amplitudes have very large uncertainties 
associated with them.
The S-wave amplitude is small at tree-level and it is no surprise 
to find the loop corrections substantially modify the leading result.
However, for the P-waves the tree-level amplitude is not small and
it would be somewhat disturbing to find large cancellations between
the higher order and tree level amplitudes.
A significant cancellation
may suggest 
a failure of chiral perturbation theory.

In the fit we find\cite{rps,walden+mjs}
\footnote{ We cannot reproduce the values of 
axial couplings (or uncertainties) found in \cite{luty+white}.}
$|C| = 1.11 \pm 0.05\ ,
D = 0.54 \pm 0.03 \ ,
F = 0.33 \pm 0.03 \ ,
{\cal H} =-1.75 \pm 0.46 \ ,
h_C = 0.30 \pm 6.75 \ ,
h_D = -0.42 \pm 0.21 \ ,$
and
$ h_F = 0.94 \pm 0.44$
when a decuplet-octet mass splitting of zero 
is used.  
The sign of $C$ is not determined as it can be changed 
by a field redefinition
and  presents no ambiguity in our results.
The errors on $h_C$ are large because the nonleptonic decays
of octet baryons  depend  on $h_C$ only through loops.  
The difference between the tree level and loop level extracted parameters
is larger than one might expect from naive dimensional analysis.
However, it is only observables that are required to be stable 
in order to have confidence
that the chiral expansion is  converging.

\section{Discussion}

The S-wave and P-wave amplitudes for the $ \Lambda\Lambda K^0 $ interaction
are explored at loop level in chiral perturbation theory. This
adds to the body of knowledge on $\Delta s = 1$
transitions in chiral perturbation theory already existing from 
the nonleptonic
hyperon decays, and the $KNN$ couplings which can be measured
in single $\Lambda$ hypernuclear decay.  Comparing these
calculations with data yields insight into the strengths and weaknesses
of the theory, and highlight which results can be considered
typical for the theory, and which anomalous.

The tree-level $\Lambda \Lambda K$ 
S-wave amplitude is small, making
it vulnerable to higher order corrections.
The S-wave amplitudes for both
nonleptonic hyperon decays and the $KNN$ couplings
are generally of order one or larger at tree level. 
The tree-level $\Lambda \Lambda K$
P-wave amplitude, on the other hand,
is substantial.  Like the P-wave nonleptonic hyperon decay amplitude,
the $\Lambda \Lambda K$
coupling has contributions from two pole diagrams.  But unlike
the nonleptonic hyperon amplitudes, a significant cancellation does
not occur.  This lends credence to the claim that the situation
with the nonleptonic hyperons may be anomalous.  
For the $KNN$
amplitudes, only one P-wave diagram contributes and no
cancellation is possible.
The errors associated with these results come from the 
propagation of errors associated with fitting other
observables at tree-level  and are
of the expected magnitude.

Looking at the one-loop results with $\Delta m =0$, we note that 
the central values are not
dramatically different from the tree level values, but 
the S-wave errors have
increased to the point where we question whether the
chiral expansion is converging for this coupling.  This is
in contrast to the situation with the nonleptonic hyperon
decays where the experimental S-waves are well reproduced at 
loop level.
The $KNN$ S-wave couplings suffer from fairly large (up to
50\%) corrections from one loop diagrams, but the associated
errors are small enough that the predictions remain meaningful.
The P-waves for the nonleptonic hyperon decays poorly
reproduce the data at one loop and experience
large corrections.  If it is true that this is caused by
the accidental cancellations that make the tree level
values small, we should not see a similar trend for
the $KNN$ or $\Lambda \Lambda K$ couplings.  Indeed, the
$KNN$ P-wave loop corrections are not large, and the
associated errors are reasonable.  The predictions for
$\Lambda \Lambda K$ with $\Delta m =0$ also look very
stable, with a small correction and small errors.
This alone might make us think that the P-wave couplings
are generally well behaved but that predictions for
processes which are anomalously small at tree level
cannot be trusted.  Once again, the errors come from
a propagation of the errors from the global parameter
fit.  They are stable under reasonable changes in
the meson decay constant, and the chiral symmetry
breaking scale.

An idea of the stability of our result to higher
order corrections can be obtained by considering the
situation when $\Delta m \ne 0$.  
The P-wave amplitude acquires a significantly increased error.
It is interesting that, despite this, the central value
remains basically unchanged.  Yet the large error
indicates that the P-wave $\Lambda \Lambda K$ coupling
could be sensitive to higher order corrections. In contrast
to the P-wave amplitudes for the nonleptonic decay
of octet baryons, where the trouble comes
from accidental cancellations between pole graphs at
tree level, here the difficulty comes because the
parameters appear in the $\Lambda \Lambda
K$ formula with nonzero $\Delta m$ in such a way that  errors
on the parameters
conspire to make the error on the result large.
The S-wave amplitude in $ \Lambda\Lambda K^0 $ suffers both from
a large uncertainty and a small tree level value.
This situation is very different from that for the
$NNK$ amplitudes, where the uncertainties even with nonzero $\Delta m$
were found to be reasonably small, 
$\sim 20\% $ or smaller.

In summary, we find that at one-loop level 
(without local counterterms which begin at 
${\cal O}(m_s)$,  formally higher order) 
the uncertainty in the $\Lambda \Lambda K$
amplitudes
appears to nearly 100\%  for 
the P-waves and 
more than 200\%  for the S-waves. 
These numbers include an estimate of the theoretical error resulting 
from the neglect of higher order terms in the chiral expansion.
The indication is that
the $\Lambda \Lambda K$ coupling may be one of the quantities
which is not well predicted by a loop level calculation in
chiral perturbation theory.  This is not the case for
other processes.
The fact that the parameters in one loop calculations fit
the known data on axial currents, the strong decays of the
decuplet, and the S-wave nonleptonic hyperon decays so well
shows that, for these processes, higher order counter terms
are not required.  

A measurement of the P-wave $\Lambda \Lambda K$ coupling
through the decay of $\Lambda \Lambda$ hypernuclei will
yield important information on chiral perturbation theory,
its parameters, and the importance of higher order
terms.  We are encouraged by the promise of experiments
at KEK and BNL, and hope that possibilities at Frascati
and TJNAF will also be pursued.

\bigskip

This was supported in part by
a grant from the Department of Energy,  DE-FG02-96ER40945. 
We thank C. Bennhold for several discussions that led
to this work.
We also thank T. Fakuda, E. Hungerford, and I. Melnikov 
for helpful communications.

\begin{figure}
\epsfxsize=10cm
\hfil\epsfbox{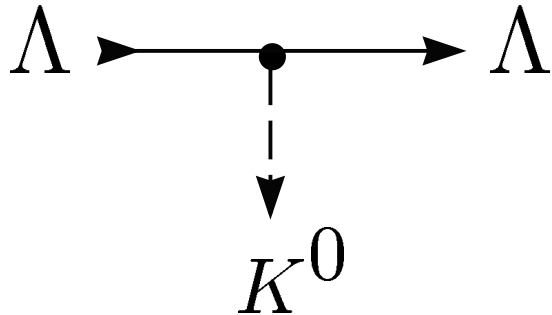}\hfill
\caption{Tree level diagram for the $\Lambda \Lambda K^0$ amplitude.  
The black dot indicates the weak vertex.}
\label{streefig}
\end{figure}

\begin{figure}
\epsfxsize=10cm
\hfil\epsfbox{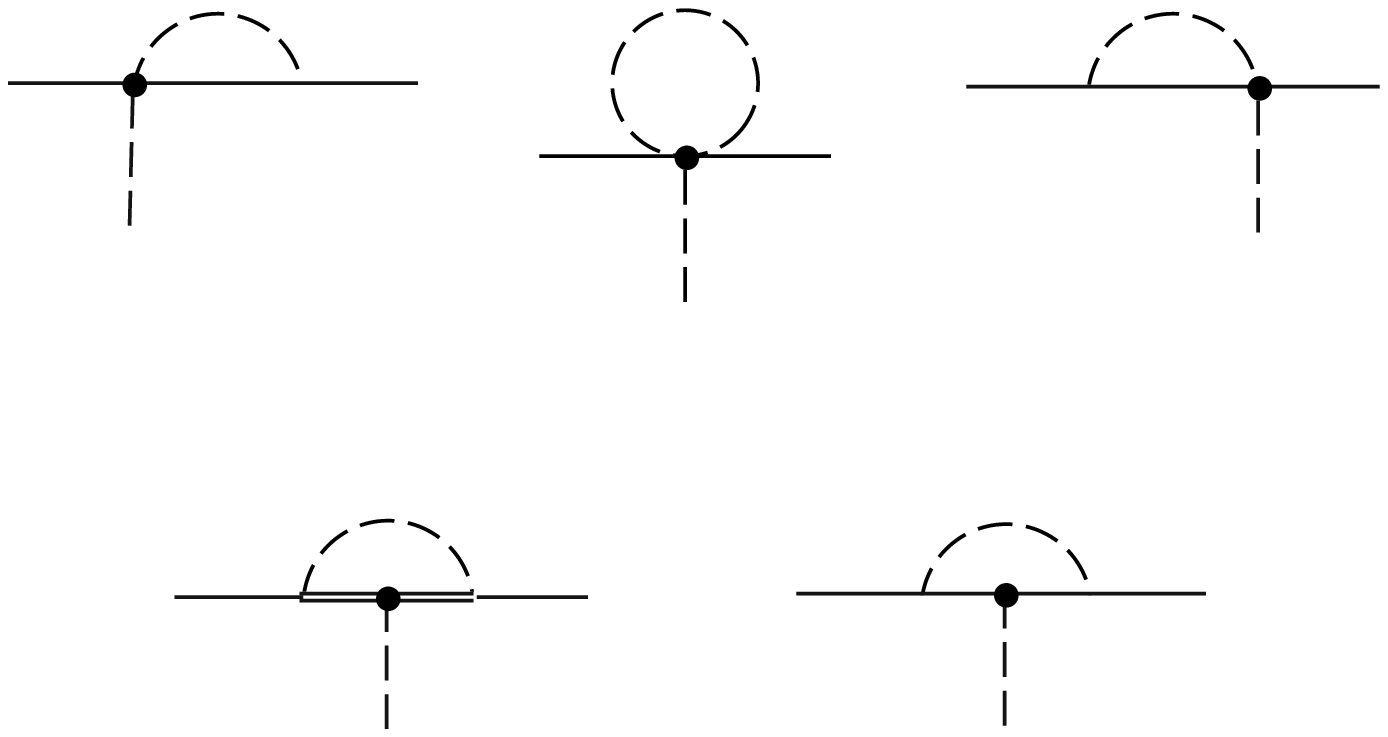}\hfill
\caption{Loop level diagrams for the S-wave amplitude.  The dashed
lines are mesons, the solid lines are octet baryons, and the
double lines are decuplet baryons.  An unmarked vertex
represents a strong interaction and the black dots are
weak interactions.}
\label{sloopfig}
\end{figure}

\begin{figure}
\epsfxsize=10cm
\hfil\epsfbox{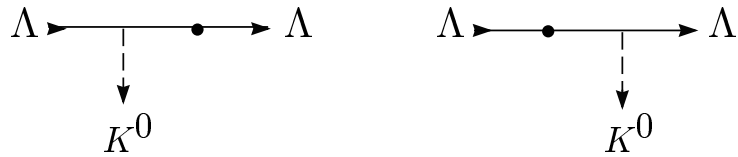}\hfill
\caption{Tree level diagrams for the $\Lambda \Lambda K^0$ amplitude.  
An unmarked
vertex represents a strong interaction and the black dots are weak 
vertices.}
\label{ptreefig}
\end{figure}

\begin{figure}
\epsfxsize=10cm
\hfil\epsfbox{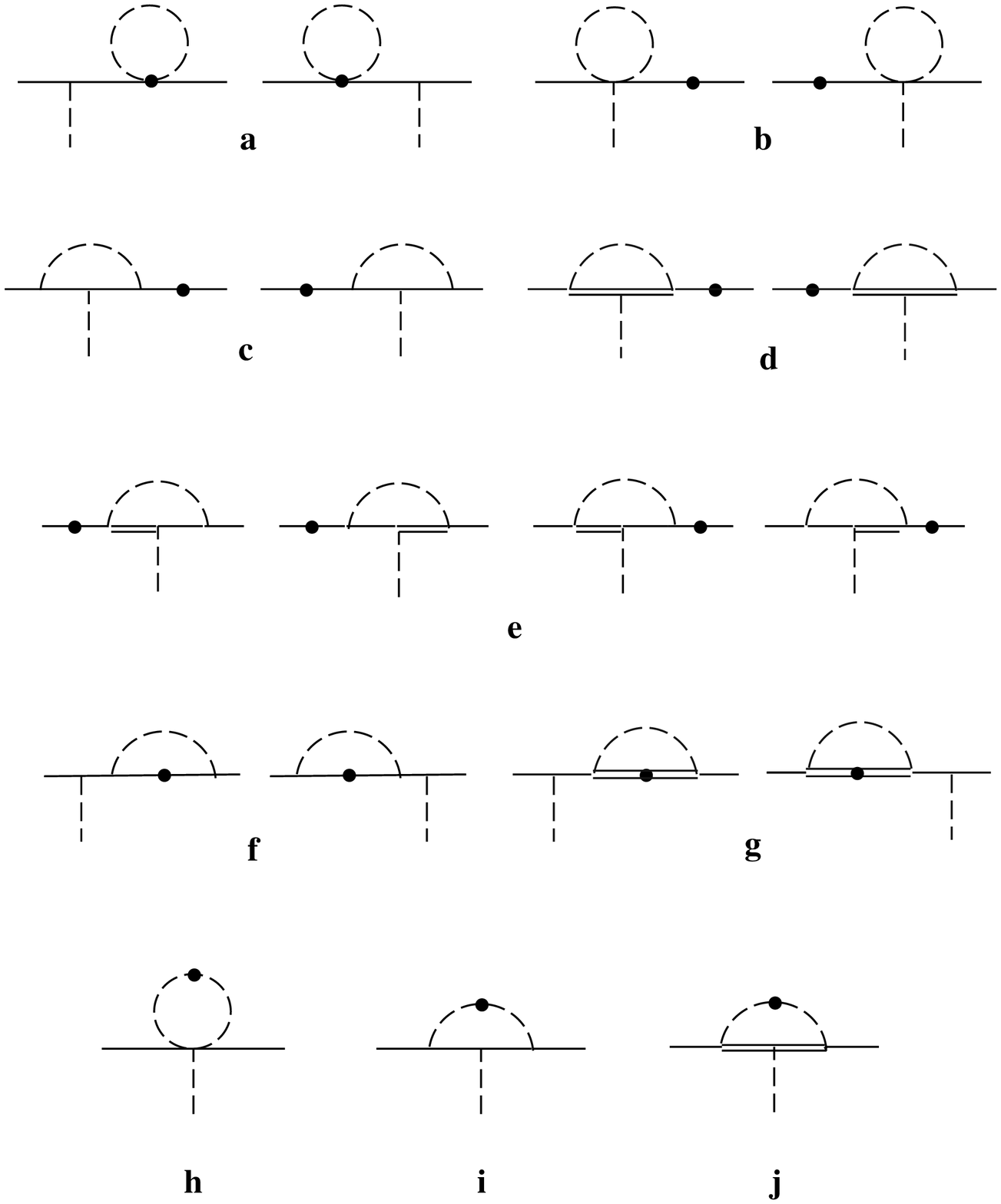}\hfill
\caption{Loop level diagrams for the P-wave amplitude.  The dashed
lines are mesons, the solid lines are octet baryons, and the
double lines are decuplet baryons.  An unmarked vertex
represents a strong interaction and the black dots are
weak interactions.}
\label{ploopfig}
\end{figure}

\begin{figure}
\epsfxsize=10cm
\hfil\epsfbox{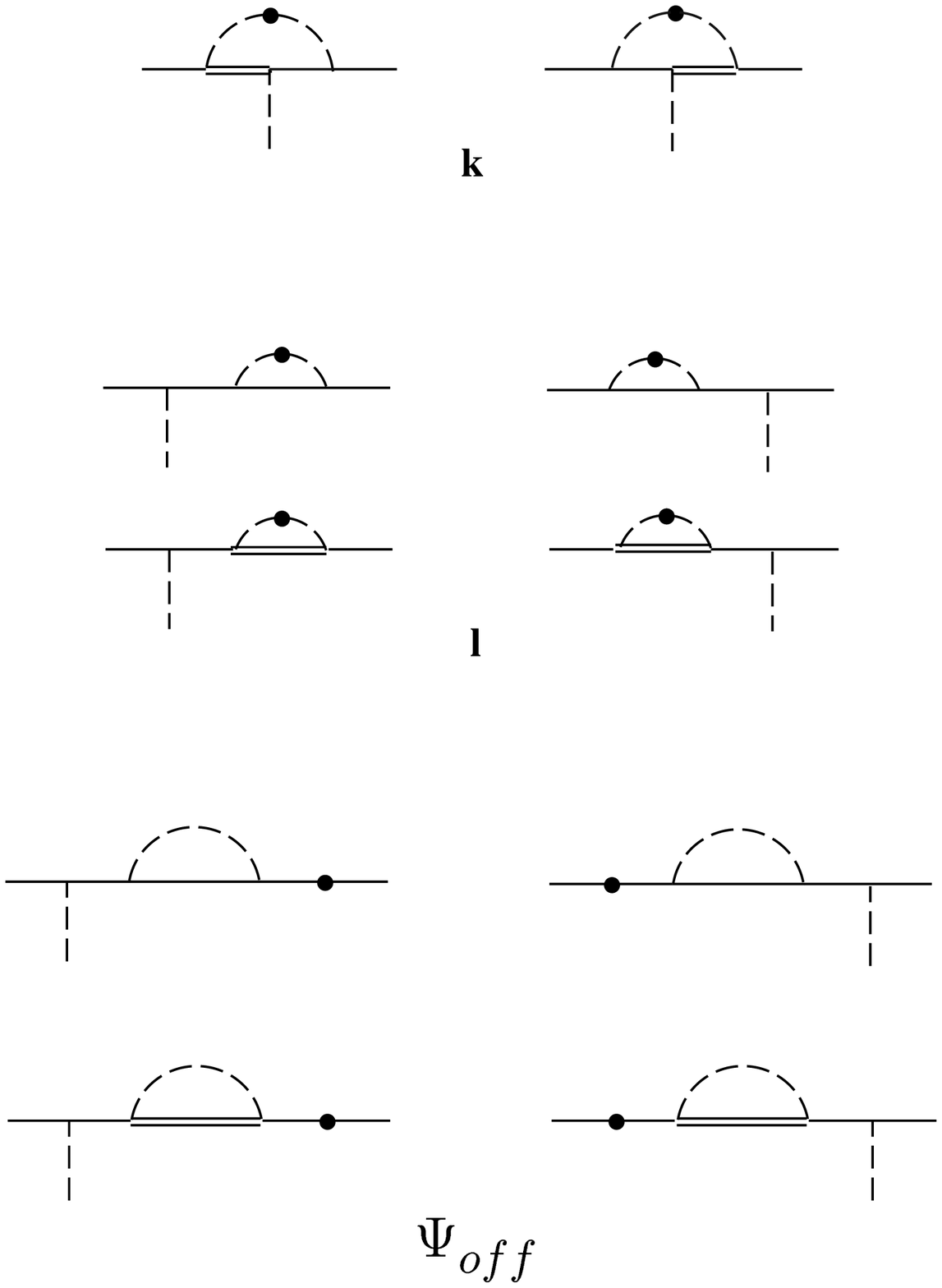}\hfill
\caption{Continuation of loop level diagrams for the P-wave amplitude.  
The dashed
lines are mesons, the solid lines are octet baryons, and the
double lines are decuplet baryons.  An unmarked vertex
represents a strong interaction and the black dots are
weak interactions. }
\label{ploop2}
\end{figure}

\end{document}